\documentclass[10pt]{article}

\usepackage{pscc}
\usepackage{graphicx}
\usepackage{epsfig,amssymb,color}
\usepackage{subfigure}
\usepackage{amsmath}

\begin{document}
\papertitle{MODELING AND CONTROL OF \\[1mm] THERMOSTATICALLY CONTROLLED LOADS}

\noindent
\names{\large Soumya Kundu}{\large Nikolai Sinitsyn}{\large Scott Backhaus}\endnames
\names{University of Michigan}{Los Alamos National Laboratory}{Los Alamos National Laboratory}\endnames
\names{Ann Arbor, USA}{Los Alamos, USA}{Los Alamos, USA}\endnames
\names{soumyak@umich.edu}{sinitsyn@lanl.gov}{backhaus@lanl.gov}\endnames

\noindent
\names{\large Ian Hiskens}\endnames
\names{University of Michigan}\endnames
\names{Ann Arbor, USA}\endnames
\names{hiskens@umich.edu}\endnames

\begin{multicols}{2}

\abstr{As the penetration of intermittent energy sources grows
  substantially, loads will be required to play an increasingly
  important role in compensating the fast time-scale fluctuations in
  generated power. Recent numerical modeling of thermostatically
  controlled loads (TCLs) has demonstrated that such load following is
  feasible, but analytical models that satisfactorily quantify the
  aggregate power consumption of a group of TCLs are desired to enable
  controller design. We develop such a model for the aggregate
  power response of a homogeneous population of TCLs to uniform
  variation of all TCL setpoints. A linearized model of the response
  is derived, and a linear quadratic regulator (LQR) has been
  designed. Using the TCL setpoint as the control input, the LQR
  enables aggregate power to track reference signals that exhibit
  step, ramp and sinusoidal variations. Although much of the work
  assumes a homogeneous population of TCLs with deterministic
  dynamics, we also propose a method for probing the dynamics of
  systems where load characteristics are not well known.}


\section{INTRODUCTION}

\PARstart{A}{s} more renewable power generation is added to power
systems, concerns for grid reliability increase due to the
intermittency and non-dispatchability associated with such sources.
Conventional power generators have difficulty in manoeuvering to
compensate for the variability in the power output from renewable
sources. On the other hand, electrical loads offer the possibility of
providing the required generation-balancing ancillary services. It is
feasible for electrical loads to compensate for energy imbalance much
more quickly than conventional generators, which are often constrained
by physical ramp rates.

A population of thermostatically controlled loads (TCLs) is well
matched to the role of load following. Research into the behavior of
TCLs began with the work of \cite{ihara} and \cite{chong}, who
proposed models to capture the hybid dynamics of each thermostat in
the population. The aggregate dynamic response of such loads was
investigated by \cite{malhame}, who derived a coupled ordinary and
partial differential equation (Fokker-Planck equation) model. The
model was derived by first assuming a homogeneous group of thermostats
(all thermostats having the same parameters), and then extended using
perturbation analysis to obtain the model for a non-homogeneous group
of thermostats. In \cite{mortensen}, a discrete-time model of the
dynamics of the temperatures of individual thermostats was derived,
assuming no external random influence. That work was later extended by
\cite{ucak} to introduce random influences and heterogeneity.

Although the traditional focus has been on direct load control methods
that directly interrupt power to all loads, recent work in
\cite{callaway} proposed hysteresis-based control by manipulating the
thermostat setpoint of all loads in the population with a common
signal. While it is difficult to keep track of the temperature and
power demands of individual loads in the population, the probability
of each load being in a given state (ON~-~drawing power or OFF~-~not
drawing any power) can be estimated rather accurately.  System
identification techniques were used in \cite{callaway} to obtain an
aggregate linear TCL model, which was then employed in a minimum
variance control law to demonstrate the load following capability of a
population of TCLs.

In this paper, we derive a transfer function relating the aggregate
response of a homogeneous group of TCLs to disturbances that are
applied uniformly to the thermostat setpoints of all TCLs. We start
from the hybrid temperature dynamics of individual thermostats in the
population, and derive the steady-state probability density functions of loads
being in the ON or OFF states. Using these probabilities we calculate
aggregate power response to a setpoint change. We linearize the
response and design a linear quadratic regulator to achieve reference
tracking by the aggregate power demand. While our analytical model
assumes a homogeneous population of loads, numerical studies are
proposed to explore situations where there is noise and heterogeneity.

\section{STEADY STATE DISTRIBUTION OF LOADS}

\subsection{Model development}

The dynamic behavior of the temperature $\theta (t)$ of a
thermostatically controlled cooling-load (TCL), in the ON and OFF
state and in the absence of noise, can be modeled by \cite{mortensen},
\begin{equation}
\dot{\theta} = \left\lbrace \begin{array}{ll} -\frac{1}{CR}\left(
  \theta - \theta_{amb} +PR \right), & \text{ON state} \\ &
  \\ -\frac{1}{CR}\left( \theta - \theta_{amb} \right), &
  \text{OFF state}	\end{array} \right. \label{micro}
\end{equation}
where $\theta_{amb}$ is the ambient temperature, $C$ is the thermal
capacitance, $R$ is the thermal resistance, and $P$ is the power drawn
by the TCL when in the ON state. This response is shown in
Figure~\ref{dynamics}.

\begin{figurehere}
\centering
\vspace{2mm}
\epsfig{file=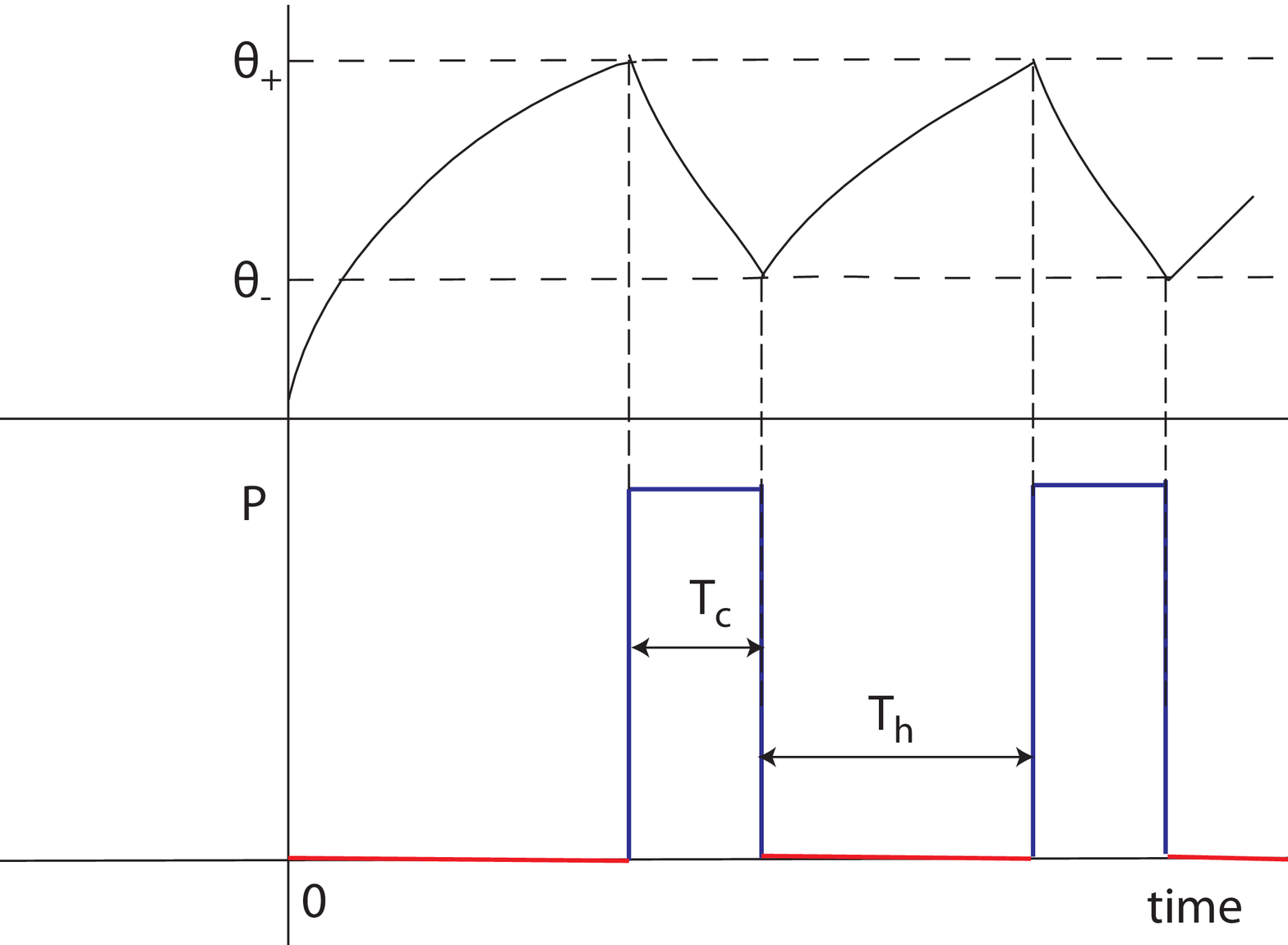, width=2.5in}
\caption{Dynamics of temperature of a thermostatic
  load.} \label{dynamics}
\vspace{2mm}
\end{figurehere}

In steady state the cooling period drives a load from temperature
$\theta _{+}$ to temperature $\theta_{-}$. Thus solving (\ref{micro})
with initial condition $\theta (0)= \theta _{+}$ gives 
\begin{multline}
\theta(t) =  \left(\theta_{amb}-PR\right) \left(1 - e^{-\frac{t}{CR}} \right) + \theta_{+} e^{-\frac{t}{CR}}. \label{theta}
\end{multline}
From (\ref{theta}) we can calculate the steady state cooling time
$T_c$ by equating $\theta (T_c)$ to $\theta_{-}$,
\begin{equation}
T_c = {CR} \ln \left( \frac{PR+\theta_{+}-\theta_{amb}}
{PR+\theta_{-}-\theta_{amb}} \right). \label{Tc}
\end{equation}
A similar calculation for the heating time gives,
\begin{equation}
T_h = {CR} \ln \left( \frac{\theta_{amb} - \theta_{-}} {\theta_{amb} -
  \theta_{+}} \right). \label{Th}
\end{equation}
In general, the expressions for the times $t_c (\theta_f)$ and $t_h
(\theta_f)$ taken to reach some intermediate temperature $\theta_f$
during the cooling and heating periods, respectively, are,
\begin{align}
t_c (\theta_f) &= {CR} \ln \left( \frac{PR+\theta_{+}-\theta_{amb}}
{PR+\theta_f-\theta_{amb}} \right) \label{tch1} \\
t_h (\theta_f) &= {CR} \ln \left( \frac{\theta_{amb} - \theta_{-}}
{\theta_{amb} - \theta_f}\right). \label{tch2}
\end{align} 

For a homogeneous\footnote{ All loads
    share the same values for parameters $\theta_{amb}$, $C$, $R$ and
    $P$.} set of TCL in steady state, the number of loads in the ON and OFF
states, $N_c$ and $N_h$ respectively, will be proportional to their
respective cooling and heating time periods $T_c$ and
$T_h$. In the absence of any appreciable noise,
  which ensures that all the loads are within the temperature
  deadband, $N_h+N_c = N$, we obtain,
\begin{align}
N_c &= \frac{T_c}{T_c + T_h} N  \\
N_h &= \frac{T_h}{T_c + T_h} N \label{NcTc}
\end{align}

By analogy, it
follows that the number of ON-loads $n_c(\theta)$ within a temperature
band of $[\theta,\; \theta_{+}]$ is proportional to the time taken $t_c(\theta)$ to cool a load down from
$\theta_{+}$ to an arbitrary temperature $\theta \geq \theta_{-}$,
\begin{align}
n_c(\theta) &= t_c(\theta) \frac{N_c}{T_c} \nonumber \\
&= t_c(\theta) \frac{N}{T_c+T_h} \label{nctc1}
\end{align}
where (\ref{NcTc}) was used to obtain (\ref{nctc1}). Likewise,
\begin{equation}
n_h(\theta) = t_h(\theta) \frac{N}{T_c+T_h}. \label{nctc}
\end{equation}
We will denote the ON probability density function by $f_1(\theta)$ and the OFF
probability density function by $f_0(\theta)$, while the corresponding cumulative
distribution functions are denoted $F_1(\theta)$ and $F_0(\theta)$,
respectively. It is to be noted that, $F_0(\theta)$ is the probability of a load being in OFF state and having a temperature $\theta \in [\theta_{-}, \theta]$ while $F_1(\theta)$ is the probability of a load being in ON state and having a temperature $\theta \in [\theta_{-}, \theta]$. Thus, $F_0(\theta)=n_h(\theta)/N$ and $F_1(\theta)=(N_c-n_c(\theta))/N$. We can therefore write,
\begin{align}
f_0(\theta) &= \frac{dF_0(\theta)}{d\theta} = \frac{d}{d\theta} \left(
\frac{n_h(\theta)}{N} \right) \nonumber \\
&= \frac{1}{N} \frac{dt_h(\theta)}{d\theta} \frac{N}{T_c+T_h} \nonumber \\
&= \frac{1}{T_c+T_h} \frac{dt_h(\theta)}{d\theta} \nonumber \\
&= \frac{CR}{(T_c+T_h)(\theta_{amb}-\theta)} \label{f0}
\end{align}
and
\begin{align}
f_1(\theta) &= \frac{dF_1(\theta)}{d\theta} = \frac{d}{d\theta} \left(
\frac{N_c-n_c(\theta)}{N} \right) \nonumber \\
&= \frac{CR}{(T_c+T_h)(PR+\theta-\theta_{amb})}. \label{f1}
\end{align}

\subsection{Simulation}

Figure~\ref{f1f0} shows a comparison of the densities calculated using
(\ref{f0}) and (\ref{f1}) and those computed from actual simulation of
the dynamics of a population of 10,000 TCLs that included a small amount of noise. The result suggests that
the assumptions underlying (\ref{f0}) and (\ref{f1}) are realistic. 
  
\begin{figurehere}
\centering
\epsfig{file=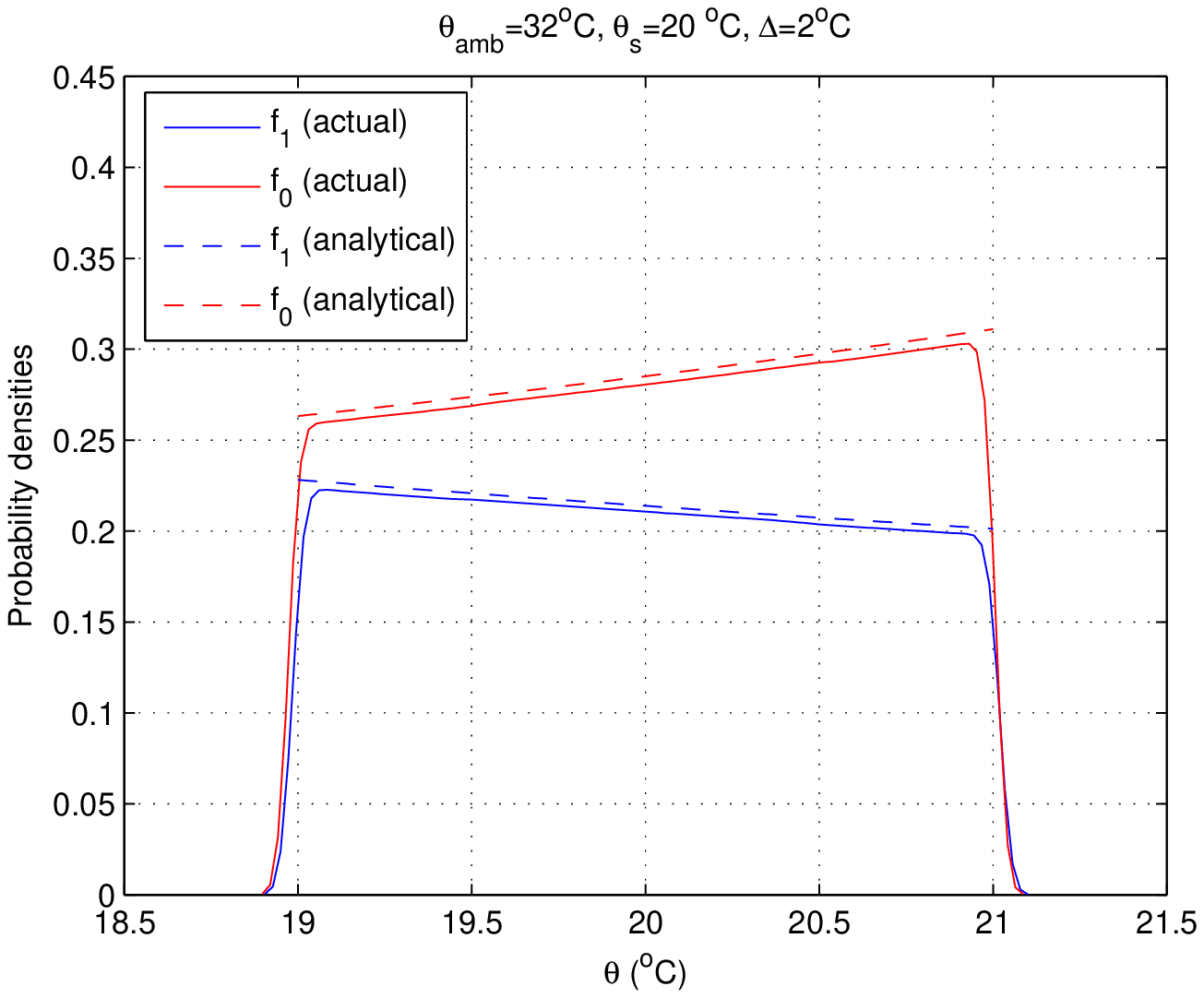, width=2.5in}
\caption{Steady state densities.}
\label{f1f0}
\end{figurehere}

\section{SETPOINT VARIATION}

Control of active power can be achieved by making a uniform adjustment
to the temperature setpoint of all loads within a large population
\cite{callaway}. It is assumed that the temperature deadband moves in
unison with the setpoint. Figure~\ref{setpoint_step} shows the change
in the aggregate power consumption of a population of TCLs for a small
step change in the setpoint of all devices. The resulting transient
variations in the OFF-state and ON-state distributions for the
population are shown in Figure~\ref{waterfall}.

\begin{figurehere}
\centering
\epsfig{file=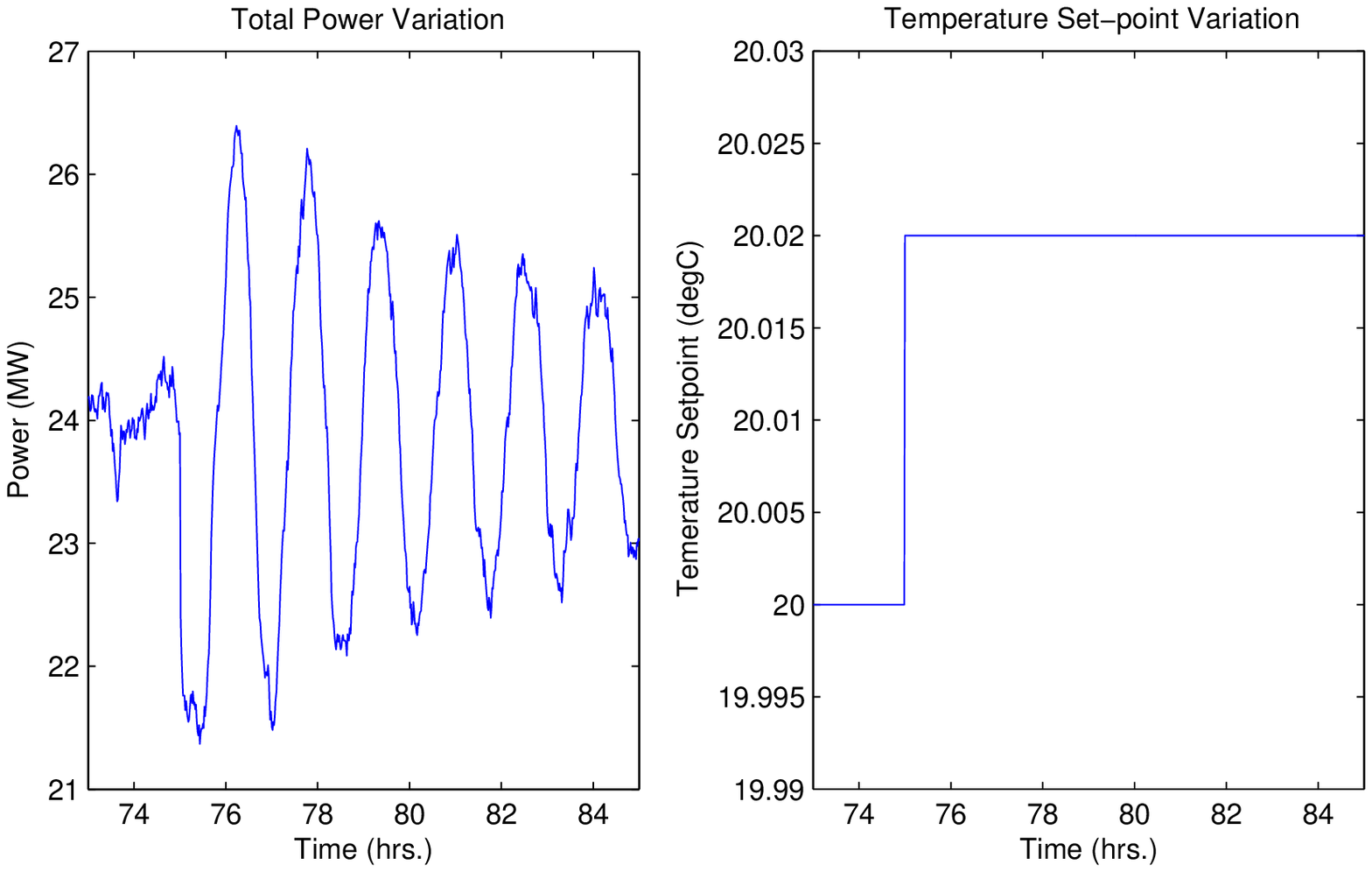, width=3in}
\caption{Change in aggregate power consumption due to a step change in
  temperature setpoint.} \label{setpoint_step} 
\end{figurehere}

\begin{figurehere}
\centering
\subfigure[OFF-state distribution.]{\epsfig{file=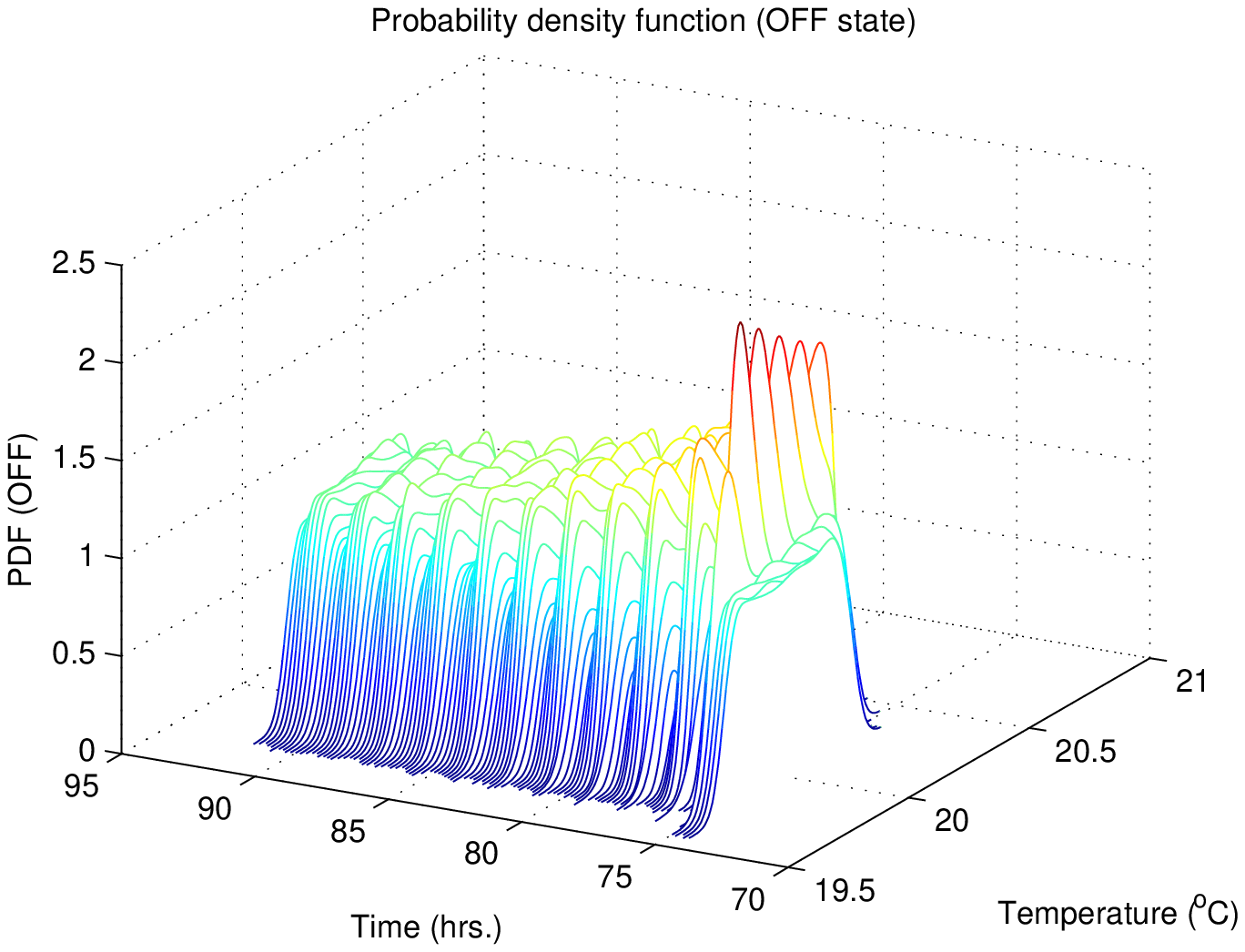, width=2.5in}\label{waterfall_OFF}}

\subfigure[ON-state distribution.]{\epsfig{file=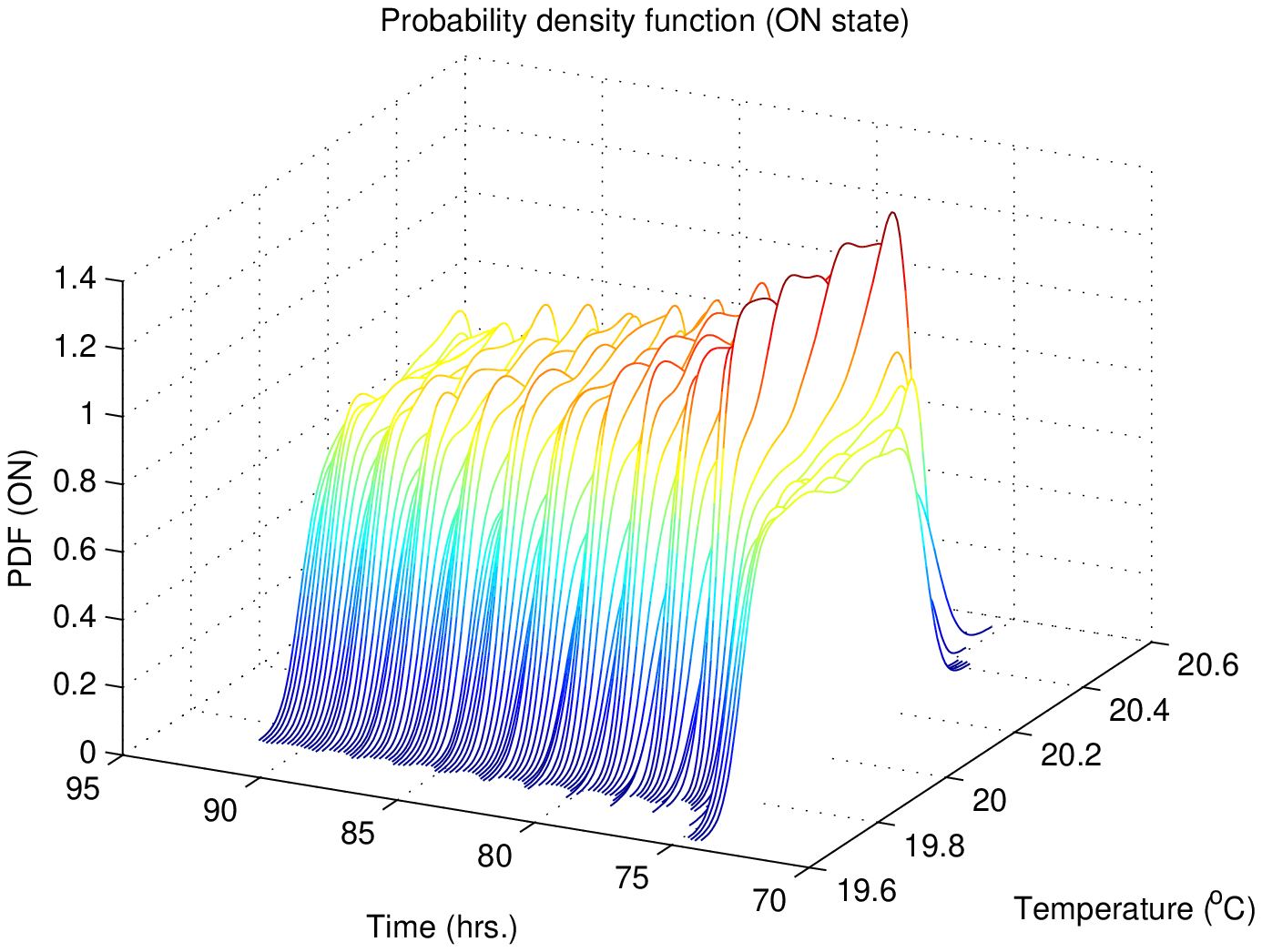, width=2.5in}\label{waterfall_ON}}
\caption{Variation in distribution of loads due to setpoint
  disturbance.} \label{waterfall}
\end{figurehere}

The aggregate power consumption at any instant in time is proportional
to the number of loads in the ON state at that instant.  The first
step in quantifying the change in power due to a step change in
setpoint is therefore to analyze the behavior of the TCL probability 
distributions. Figure~\ref{pdf_shift} depicts a situation where the
setpoint has just been increased. The original deadband ranged from
$\theta_-^0$ to $\theta_+^0$, with the setpoint at
$(\theta_-^0+\theta_+^0)/2$. After the positive step change, the new
deadband lies between $\theta_{-}$ to $\theta_{+}$, with the deadband
width $\Delta=\theta_+^0-\theta_-^0=\theta_+-\theta_-$ remaining
unchanged. The setpoint is shifted by
$\delta=\theta_{-}-\theta_-^0=\theta_{+}-\theta_+^0$. To solve for the power consumption, we need to consider four different TCL starting conditions immediately after the step change in setpoint, i.e. $a$-$d$ in Figure~\ref{pdf_shift}. Using Laplace transforms, we compute the time dependence of the power consumption for each of these loads (shown in Figure~\ref{gabcd}) and then compute the total power consumption by integrating over the distributions $f_0$ and $f_1$. At the instant
the step change in applied, the temperatures of loads at points $a$, $b$, $c$ and $d$ are
$\theta_a,\theta_b,\theta_c$ and $\theta_d$, respectively.

\begin{figurehere}
\centering
\epsfig{file=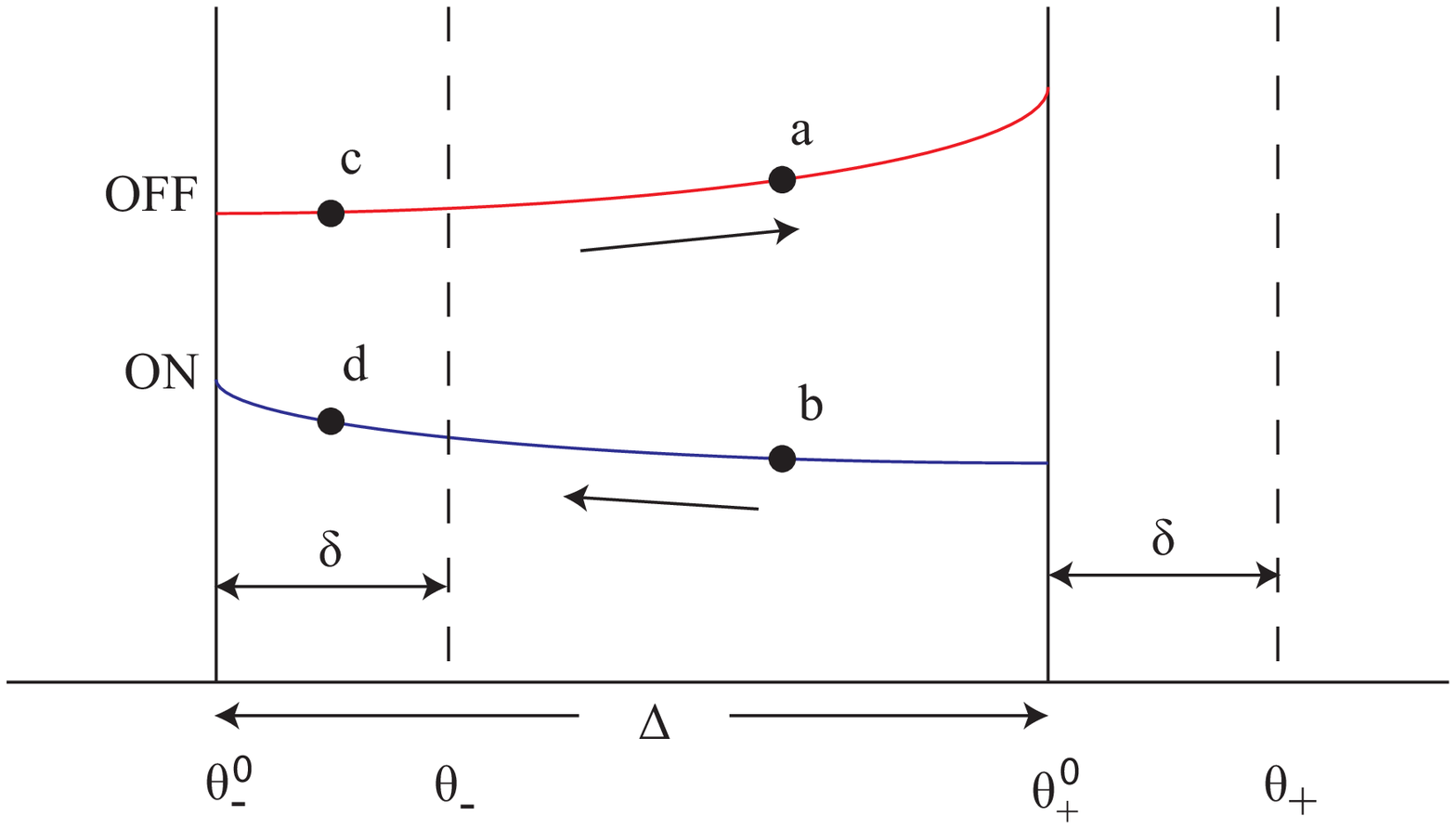, width=2.5in}
\caption{Different points of interest on the density curves.} \label{pdf_shift}
\end{figurehere}

\begin{figurehere}
\centering
{
\subfigure[Power waveform at point $a$.]{\epsfig{file=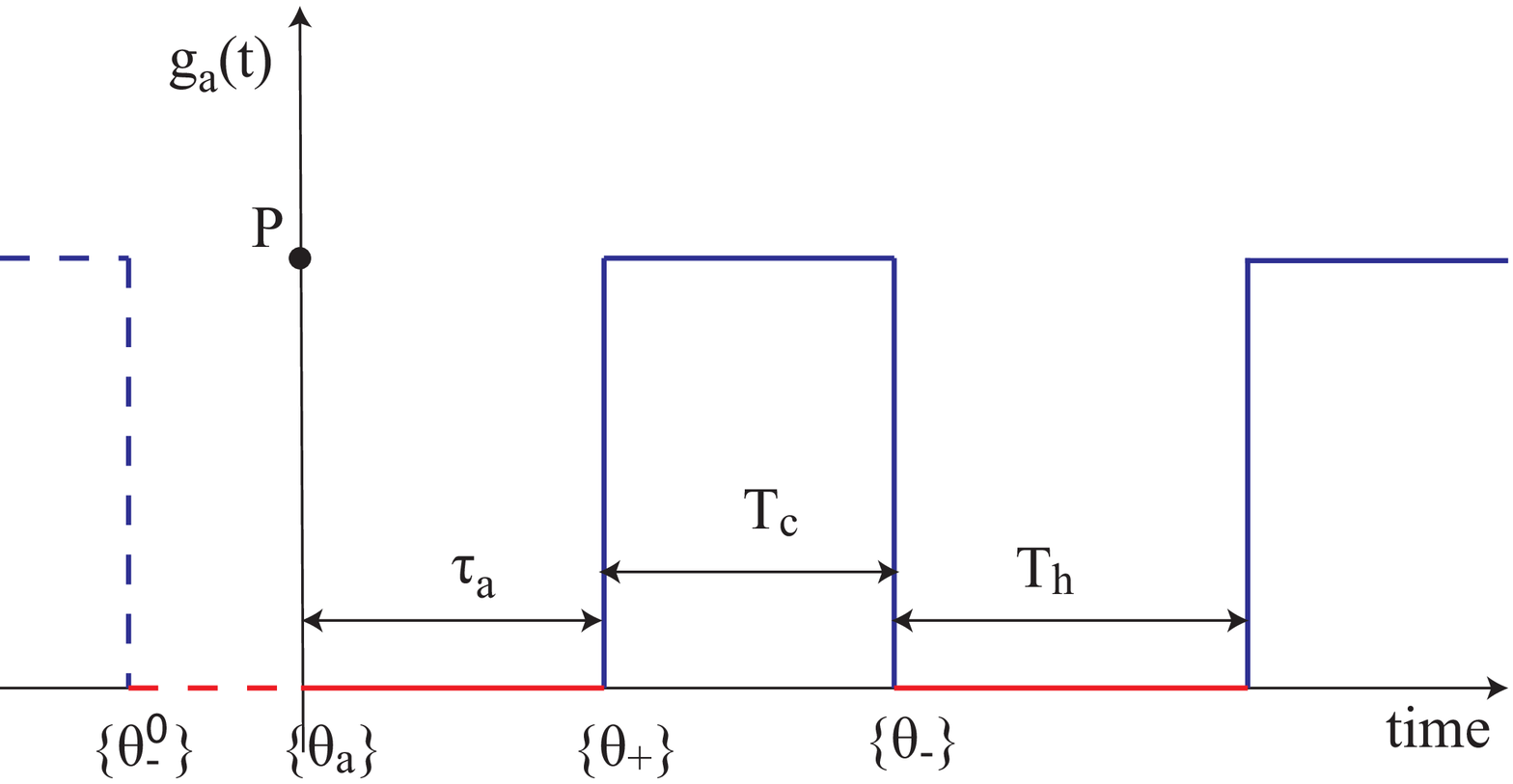, width=2.5in}\label{gas}}
\subfigure[Power waveform at point $b$.]{\epsfig{file=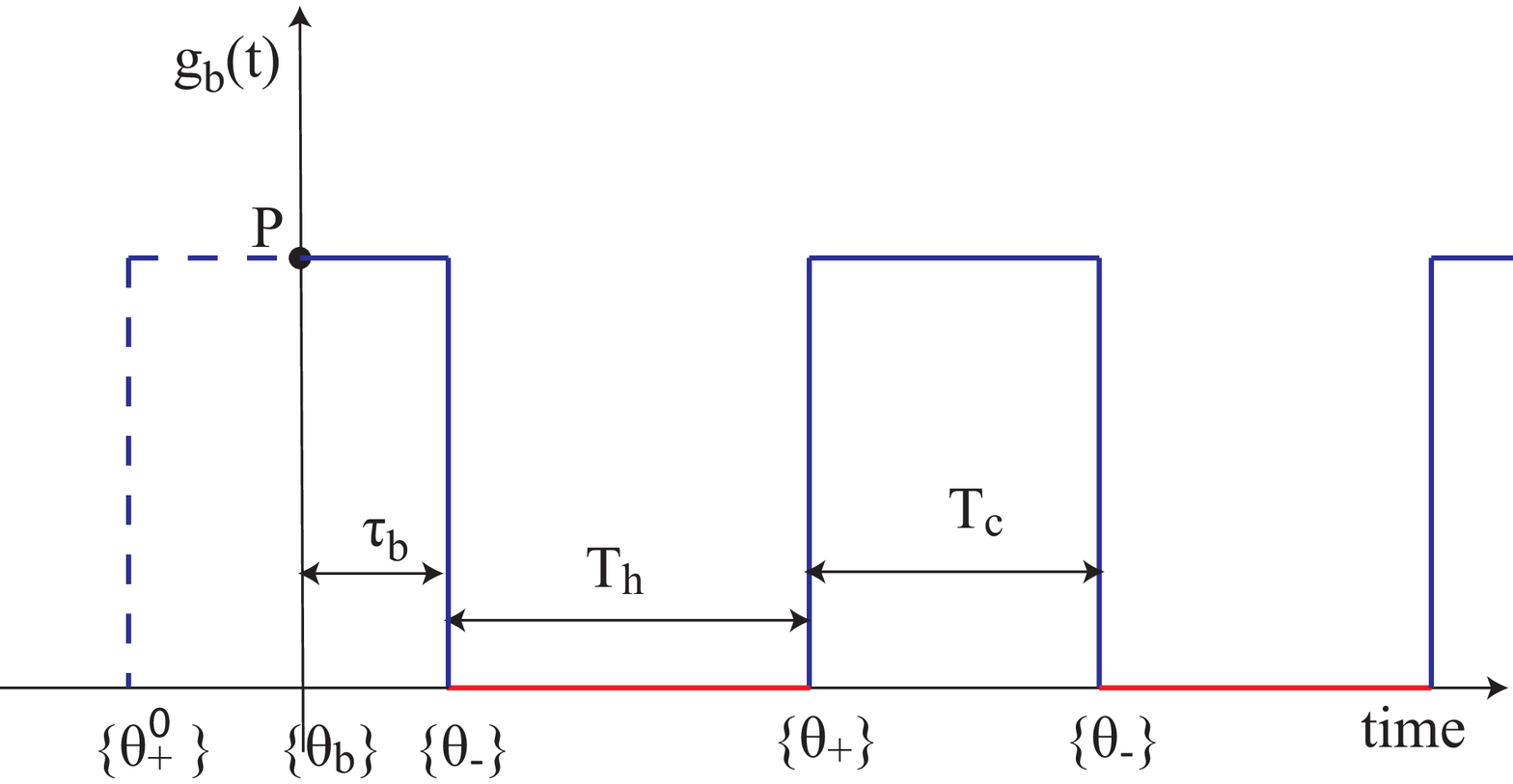, width=2.5in}\label{gbs}}
\subfigure[Power waveform at point $c$.]{\epsfig{file=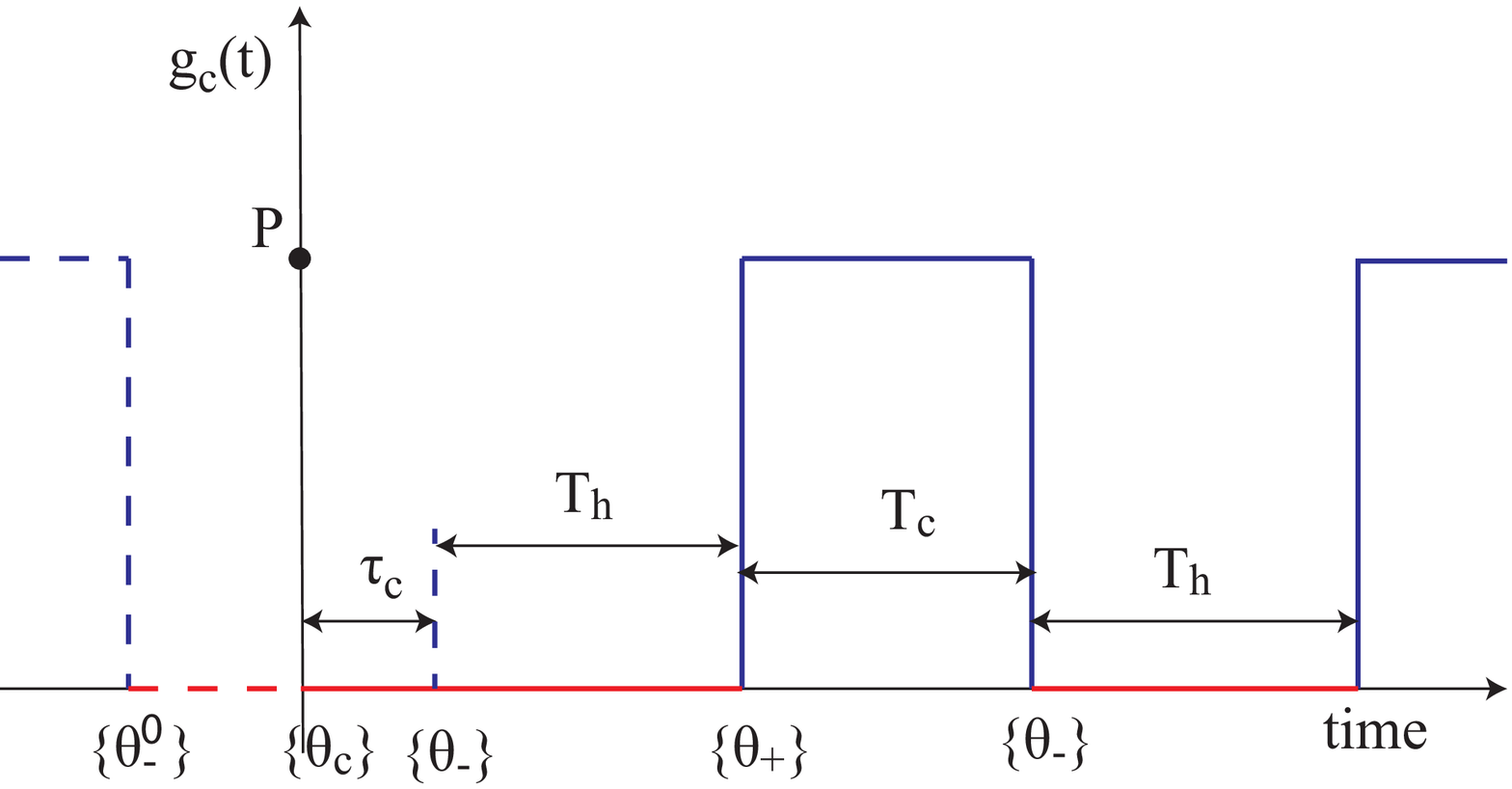, width=2.5in}\label{gcs}}
\subfigure[Power waveform at point $d$.]{\epsfig{file=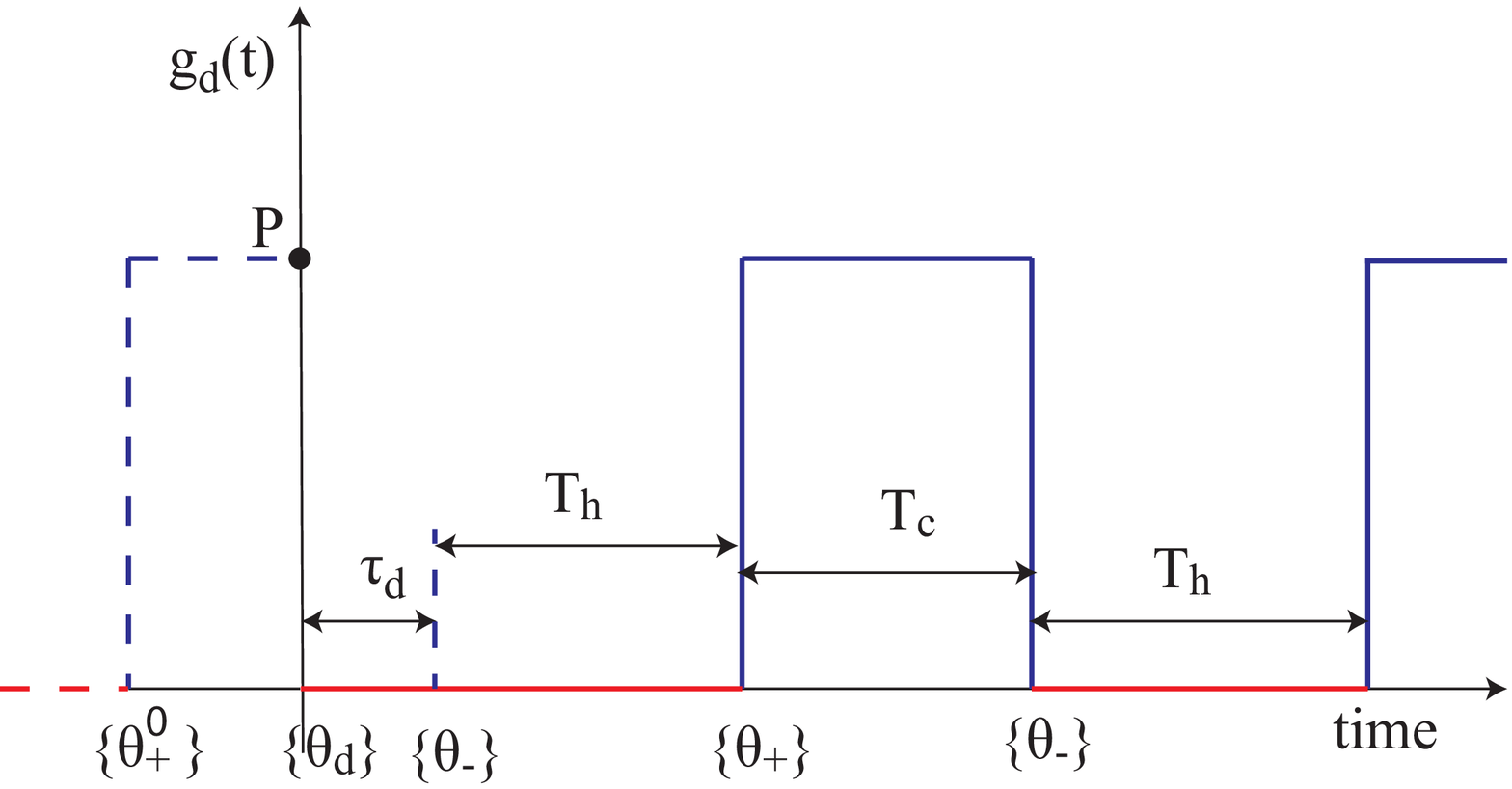, width=2.5in}\label{gds}}
\caption{Power waveforms at four different points marked in
  Figure~\ref{pdf_shift}.}}\label{gabcd}
\end{figurehere}

The power consumption $g_a(t,\tau_a)$ of the load at $a$ starting from
the instant when the step change in setpoint is applied is shown in
Figure~\ref{gas}. All the loads in the OFF-state and having a
temperature between $\theta_-$ and $\theta_+^0$ at the instant when
the deadband shift occurs will have power waveforms similar in nature
to $g_a(t,\tau_a)$. Thus the load at $a$ typifies the behavior of all
the loads lying on the OFF-state density curve between $\theta_{-}$
and $\theta_+^0$. The same argument applies for loads at points $b$,
$c$ and $d$. Figures~\ref{gas}-\ref{gds} illustrate the general nature
of the power waveforms of the loads in all four regions, marked by
$a$, $b$, $c$ and $d$ in Figure~\ref{pdf_shift}.

The Laplace transform of $g_a(t,\tau_a)$ is
\[
\mathbf{G}_a(s,\tau_a)=e^{-s\tau_a} \mathbf{G}(s)
\]
where
\[
\mathbf{G}(s)=\frac{P(1-e^{-sT_c})}{s(1-e^{-s(T_c+T_h)})}
\]
and $\tau_a=T_h-t_h(\theta_a)$, with $t_h(\theta_a)$ given by
(\ref{tch2}). Averaging over all such loads (represented by $a$) on the
OFF density curve between temperatures $\theta_{-}$ and $\theta_+^0$,
we obtain the Laplace transform of the average power demand,
\begin{equation}
\mathbf{P}_a(s) = \int_{\theta_{-}}^{\theta_+^0}{ f_0(\theta_a)
  \mathbf{G}_a(s,\tau_a)} d\theta_a \label{pas}
\end{equation} 
where $f_0(\theta_a)$ can be computed from (\ref{f0}).

In Figure~\ref{gbs}, a load at point $b$ on the ON density curve in
Figure~\ref{pdf_shift} has power consumption $g_b(t,\tau_b)$, where
$\tau_b=T_c-t_c(\theta_b)$, and $t_c(\theta_b)$ is given by
(\ref{tch1}). The Laplace transform is,
\[
\mathbf{G}_b(s,\tau_b) = \Big( e^{s (T_c-\tau_b )}
\mathbf{G}(s) - \frac{P}{s}
\big(e^{s\left(T_c-\tau_b\right)}-1 \big) \Big).
\]
We can compute the average power demand of all the loads represented
by $b$ as
\begin{equation}
\mathbf{P}_b(s) = \int_{\theta_-}^{\theta_+^0} f_1(\theta_b)
\mathbf{G}_b (s,\tau_b) d\theta_b.  \label{pbs}
\end{equation}

In Figure~\ref{gcs}, the power consumption $g_c(t,\tau_c)$ of a load
at point $c$ on the OFF density curve in Figure~\ref{pdf_shift} has the
Laplace transform
\[
\mathbf{G}_c(s,\tau_c) = e^{-s(T_h+\tau_c)} \mathbf{G}(s),
\]
where $\tau_c = CR \ln
\left(\frac{\theta_{amb}-\theta_-}{\theta_{amb}-\theta_c}\right)$. The average power demand of the loads represented by
  the point $c$ is then given by
\begin{equation}
\mathbf{P}_c(s) = \int_{\theta_-^0}^{\theta_-} f_0(\theta_c)
\mathbf{G}_c(s,\tau_c) d\theta_c \label{pcs}
\end{equation}

Figure~\ref{gds} depicts the situation of a load at point $d$ on the
ON density curve, that suddenly switches to the OFF state as the
deadband is shifted (for now we assume the deadband is shifted to the
right, i.e., there is an increase in the setpoint). The power consumption
$g_d(t)$ has the Laplace transform
\[
\mathbf{G}_d(s,\tau_d) = e^{-s(T_h+\tau_d)} \mathbf{G}(s),
\]
where the dynamics in (\ref{micro}) can be solved for
$\tau_d = CR \ln \left(
\frac{\theta_{amb}-\theta_d}{\theta_{amb}-\theta_-} \right)$. The average power demand of the loads characterized
by point $d$ in Figure~\ref{pdf_shift} is then given by
\begin{equation}
\mathbf{P}_d(s) = \int_{\theta_-^0}^{\theta_{-}} f_1(\theta_d)
\mathbf{G}_d (s,\tau_d) d\theta_d. \label{pds}
\end{equation}

The average power demand of the whole population becomes,
\begin{equation}
\mathbf{P}_{avg}(s) = \mathbf{P}_a(s) + \mathbf{P}_b(s) +
\mathbf{P}_c(s) + \mathbf{P}_d(s).
\end{equation}
Using (\ref{pas}), (\ref{pbs}), (\ref{pcs}) and (\ref{pds}) we obtain
an expression for $\mathbf{P}_{avg}(s)$ that is rather complex. It is
hard, and perhaps even impossible, to obtain the inverse Laplace
transform. However, with the assistance of MATHEMATICA$^\circledR$,
$\mathbf{P}_{avg}(s)$ may be expanded as a series in $s$. We also make
use of the assumptions,
\begin{gather*}
\Delta \ll (\theta_s- \theta_{amb} + PR) \\
\Delta \ll (\theta_{amb} - \theta_s) \\
\delta \ll \Delta
\end{gather*}
where $\theta_s$ is the setpoint temperature. Note that the first two
assumptions require that the deadband width is small, while the third assumption requires that the shift in the deadband
  is small relative to the deadband width. This latter assumption
  ensures that the load densities are not perturbed far from their
  steady-state forms. Accordingly, the steady-state power consumption
is given by
\[
P_{avg,ss} \approx \frac{(\theta_{amb}-\theta_+)N}{\eta R},
\]
where $\eta$ is the electrical efficiency of the cooling equipment and
$N$ is the population size. The deviation in power response can be
approximated by
\begin{equation}
\mathbf{P}_{tot}(s) \approx - \left( \frac{d}{s} + \frac{\omega
  A_{\Delta}} {s^2+\omega^2}\right) \delta \label{ptots} 
\end{equation}
where
\begin{gather*}
A_{\Delta} = \frac{5 \sqrt{15} C (\theta_{amb}-\theta_+)
  (PR-\theta_{amb}+\theta_+) }{\eta \big(
  P^2R^2 + 3PR(\theta_{amb}-\theta_+) - 3(\theta_{amb}-\theta_+)^2
  \big)^{3/2} }  \\
\qquad\qquad\qquad\times \frac{(3PR-\theta_{amb}+\theta_+)
  N}{(T_{c0}+T_{h0})}, \\
\omega = \frac{2 \sqrt{15} (\theta_{amb}-\theta_+) (PR-\theta_{amb}+\theta_+)}
       {CR\Delta \sqrt{P^2R^2 + 3PR(\theta_{amb}-\theta_+) -
           3(\theta_{amb}-\theta_+)^2}}, \\
d = \frac{N}{\eta R}.
\end{gather*}
and $T_{c0}$ and $T_{h0}$ are the original (prior to the setpoint
shift) steady-state cooling and heating times, respectively, given by
(\ref{Tc}) and (\ref{Th}). The transfer function for this linear model
is,
\[
\mathbf{T}(s) = \frac{\mathbf{P}_{tot}(s)}{\delta/s} = -\Big( d +
\frac{A_{\Delta} \omega s}{s^2+\omega^2} \Big).
\]

Due to the assumptions of low-noise and homogeneity, our analytical
model is undamped. The actual system, on the other hand, experiences
both heterogeneity and noise, and therefore will exhibit a damped
response. In order to capture that effect, we have chosen to add a
damping term $\sigma$ (to be estimated on-line) into the model, giving
\begin{equation}
\mathbf{T}(s) = -\Big(d + \frac{s \omega A_{\Delta}}{(s+\sigma
)^2 + \omega^2} \Big). \label{shift_tf}
\end{equation}

Figure~\ref{shift_check} shows a comparison between the response
calculated from the model (\ref{shift_tf}) and the true response to a
step change in the setpoint obtained from simulation.  A damping
coefficient of 0.002~min$^{-1}$ was added, as that value gave a close
match to the decay in the actual system response.

\begin{figurehere}
\centering
\epsfig{file=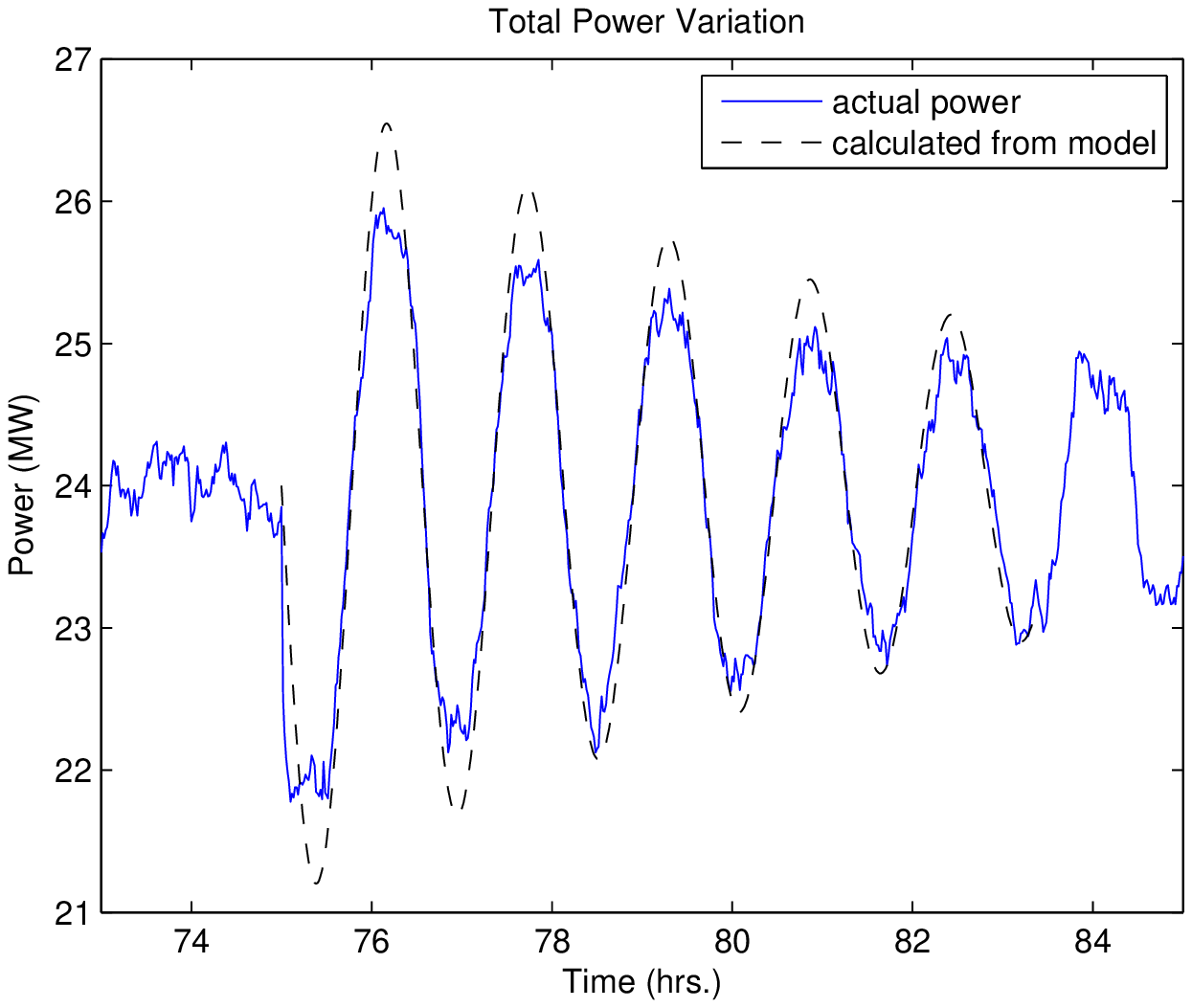, width=2.5in}\label{shift_check}
\caption{Comparison of the approximate model with the actual
  simulation, for the same setpoint disturbance as in
  Figure~\ref{setpoint_step}.}
\end{figurehere}

\section{CONTROL LAW}

The TCL load controller, described by the transfer function
(\ref{shift_tf}), can also be expressed in state-space form,
\begin{align*}
\dot{x} &= \mathbf{A} x+\mathbf{B} u \\
y &= \mathbf{C} x+\mathbf{D} u
\end{align*}
where the input $u(t)$ is the shift in the deadband of all TCLs, and
the output $y(t)$ is the change in the total power demand from the
steady-state value. The state-space matrices are given by
\begin{align*}
\mathbf{A} &= \begin{bmatrix} -2\sigma & -\omega
  \\ \frac{\sigma^2+\omega^2}{\omega} & 0 \end{bmatrix}, &
\mathbf{B} & = \begin{bmatrix} \omega A_\Delta
  \\ 0 \end{bmatrix}, \\
\mathbf{C} & = \begin{bmatrix} -1 & 0 \end{bmatrix},  & \mathbf{D} &= -d.
\end{align*}

Our goal is to design a controller using the linear quadratic
regulator (LQR) approach \cite{lqr} to track an exogenous reference
$y_d$. We observe that the system has an open-loop zero very close to
the imaginary axis ($d\ll \omega A_\Delta$) and hence we need to use
an integral controller. Considering the integral of the output error
$e=(y-y_d)$, where $y_d$ is the reference, as the third state $w(t) =
\int_{0}^{t} (y(\tau) - y_d(\tau) ) d\tau$ of the system, the modified
state-space model becomes
\begin{align*}
\dot{\underline{x}} &= {\bf \underline{A}}\,\underline{x}+{\bf
  \underline{B}}u+{\bf E}\,y_d \nonumber \\
y &= {\bf \underline{C}}\,\underline{x}+{\bf \underline{D}}u
\end{align*}
where $\underline{x}=[x \quad w]^\top$ and,
\begin{align*}
{\bf \underline{A}} &= \begin{bmatrix} {\bf A} & {\bf 0}_{2\times1}
  \\ {\bf C} & 0 \end{bmatrix}, & {\bf \underline{B}}
&= \begin{bmatrix} {\bf B} \\ {\bf D} \end{bmatrix}, \\
{\bf \underline{C}} &= \begin{bmatrix} {\bf C} & 0 \end{bmatrix}, &
{\bf \underline{D}} &= {\bf D}, \quad {\bf E} = \begin{bmatrix} {\bf
    0}_{2\times1}\\ -1 \end{bmatrix}.
\end{align*}

Minimizing the cost function 
\[
J = \int^{\infty}_{0} \big( \underline{x}(t)^\top Q \underline{x}(t)
+ u(t)^2 R \big) dt
\]
where $Q\geq{\bf 0}_{3\times3}$ and $R>0$ are design variables, we
obtain the optimal control law $u(t)$ of the form
\[
u = -( \mathbf{K} \, \underline{x} + \mathbf{G} \, y_d),
\]
with $\mathbf{G}$ a pre-compensator gain chosen to ensure unity DC
gain. Since we can only measure the output $y(t)$ and the third state
$w(t)$, the other two states are estimated using a linear quadratic
estimator \cite{lqr} which has the state-space form,
\begin{align*}
\dot{\hat{x}} &= \mathbf{A} \, \hat{x} + \mathbf{B} u + \mathbf{L}
(y-y_d) \\ 
\hat{y} &= \mathbf{C} \, \hat{x} + \mathbf{D} u \\
u &= - \mathbf{K} \, \begin{bmatrix} \hat{x} \\ w \end{bmatrix} +
\mathbf{G} \, y_d.
\end{align*}

\begin{figurehere}
\centering
{
\subfigure[Response to step reference and the control input]{\epsfig{file=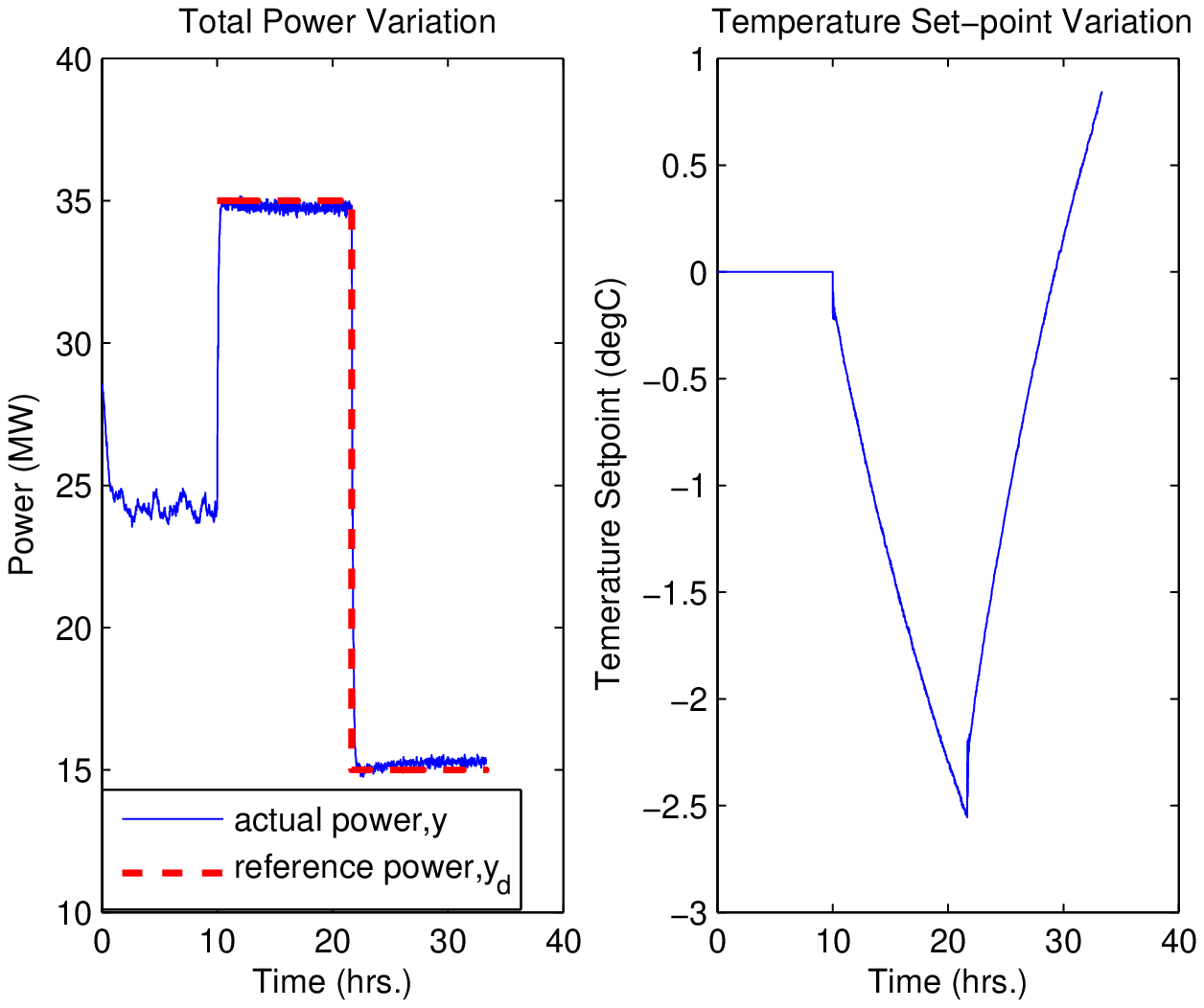, width=3in}}

\subfigure[Response to ramp reference and the control input]{\epsfig{file=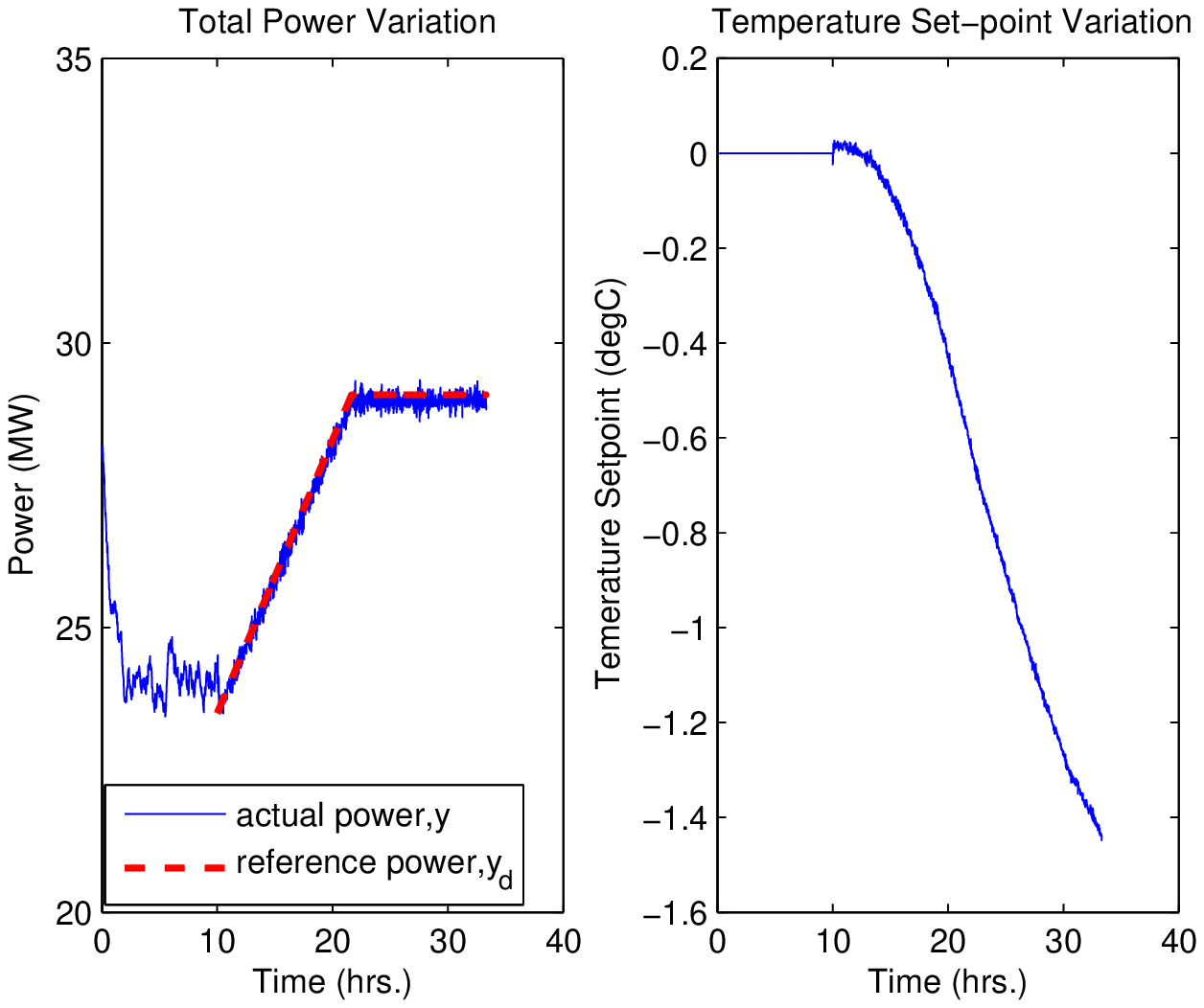, width=3in}}

\subfigure[Response to sinusoidal reference and the control input]{\epsfig{file=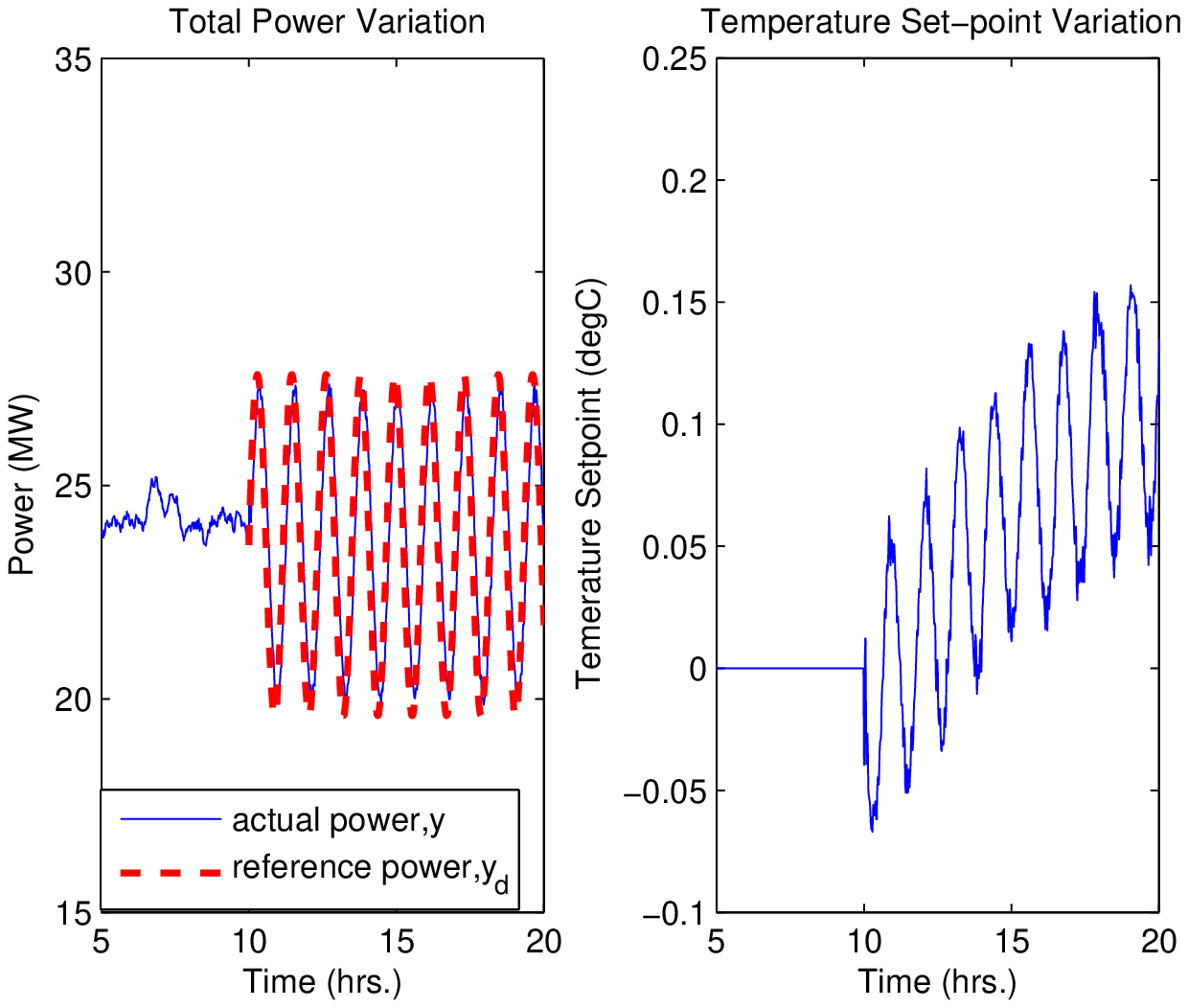, width=3in}\label{shift_sinusoid}}
\caption{Reference tracking achieved through setpoint shift}
\label{shift_response}}
\end{figurehere}

\begin{figurehere}
\centering
\subfigure[OFF-state distribution.]{\epsfig{file=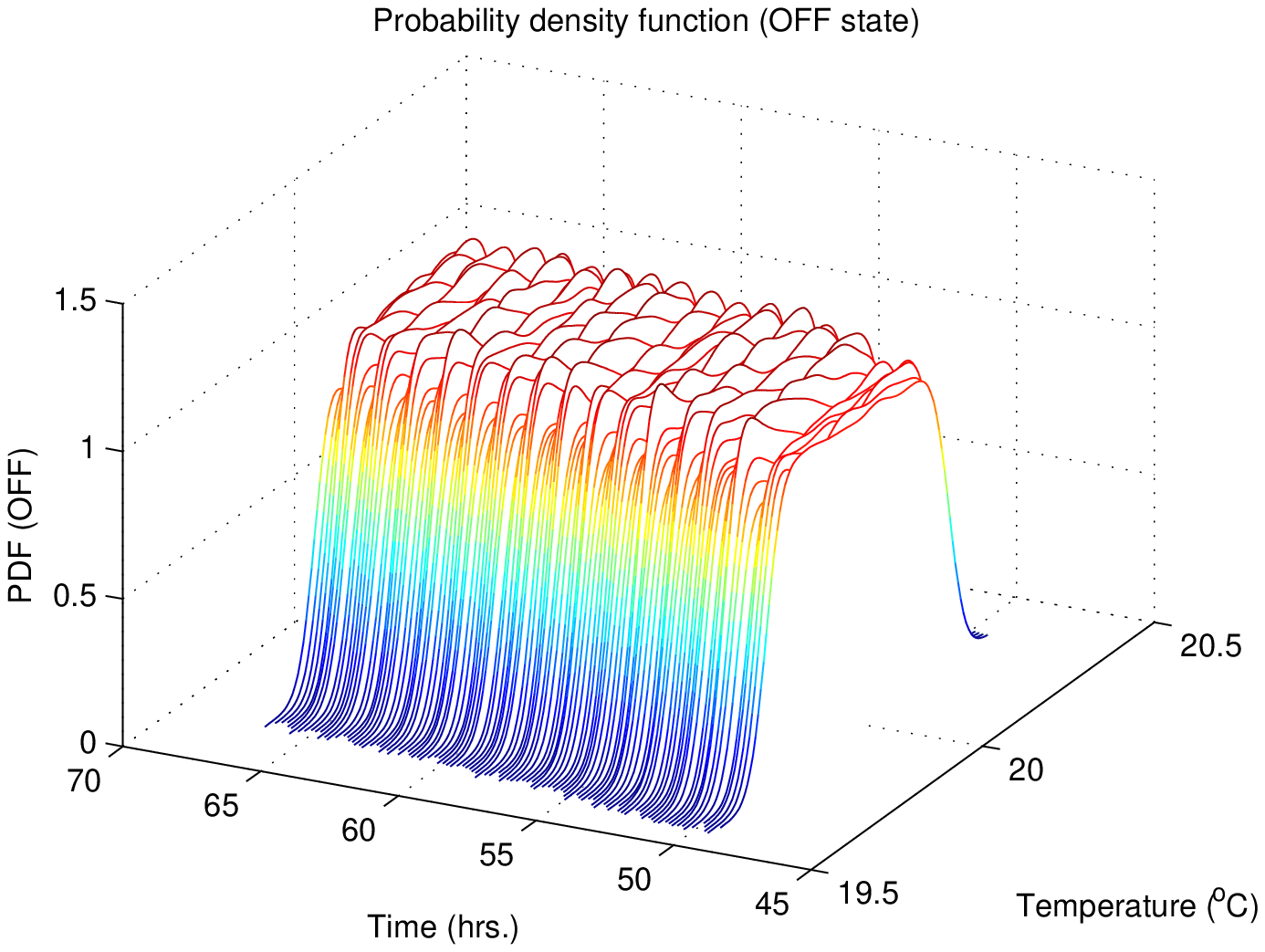, width=2.5in}}

\subfigure[ON-state distribution.]{\epsfig{file=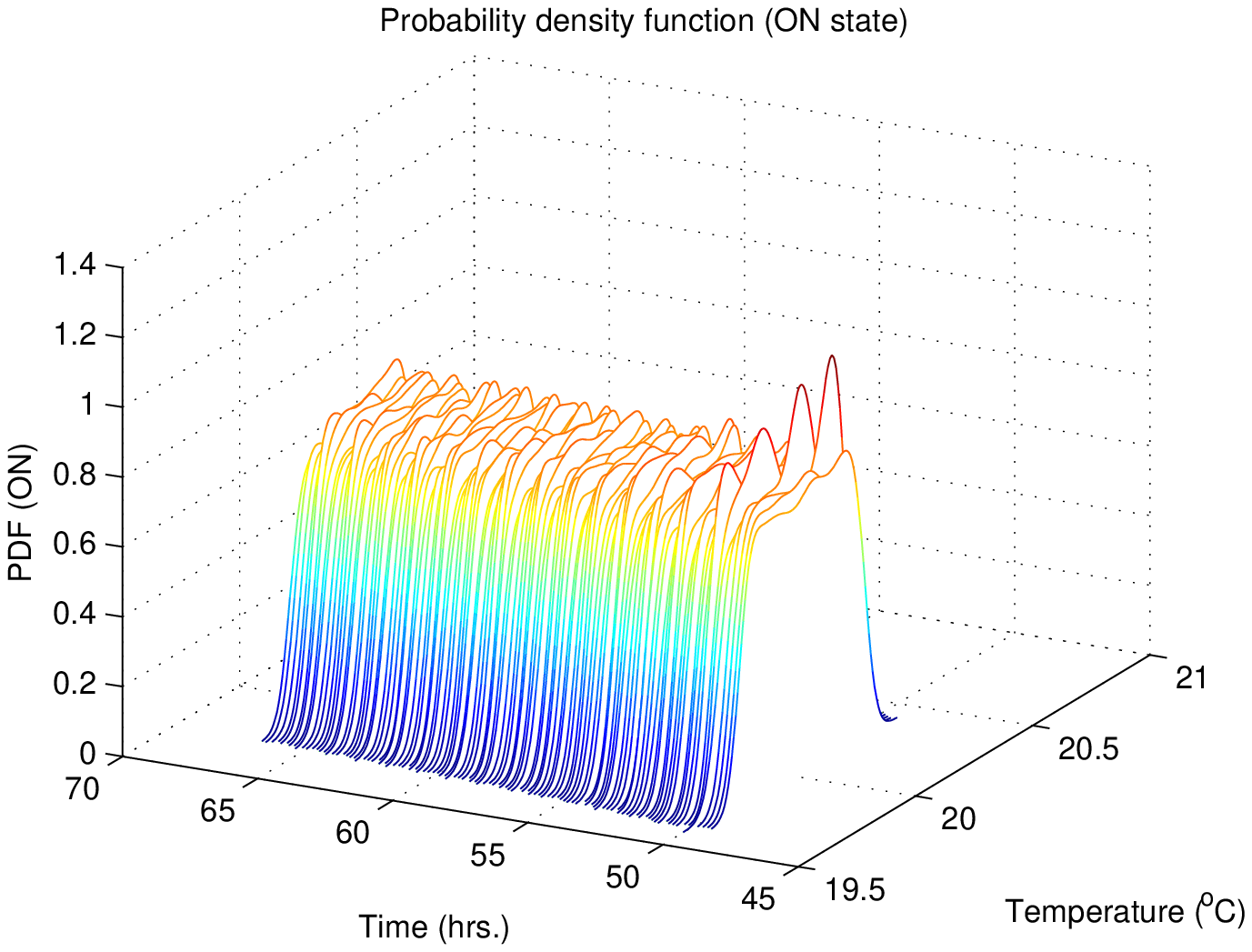, width=2.5in}}
\caption{Variation in distribution of loads under the influence of the
  controller.} \label{waterfall_CL}
\end{figurehere}

The plots in Figure~\ref{shift_response} show that the controller can
be used to force the aggregate power demand of the TCL population to
track a range of reference signals. The transient variations in the
ON-state and OFF-state populations are shown in
Figure~\ref{waterfall_CL}. In comparison with the uncontrolled
response of Figure~\ref{waterfall}, it can be seen that the controller
suppresses the lengthy oscillations. Figure~\ref{waterfall_CL} shows
that in presence of the controller, the distribution of loads {\it
  almost} always remains close to steady state, justifying an
assumption made during the derivation of the model.

\section{HETEROGENEITY AND NOISE}

The work presented in previous sections assumes a homogeneous
population of loads with deterministic dynamics. The analysis remains
valid if we consider the possibility of grouping a large number loads
having closely matched parameters, and if there is very low noise in
the system. When such assumptions no longer remain valid, we cannot
design a tracking controller based on the developed model. In such
cases, however, we propose a probing method that can balance over- or
under-production of energy over a certain duration of time.

\begin{figurehere}
\centering
\subfigure[Power response to short duration pulses.]{\epsfig{file=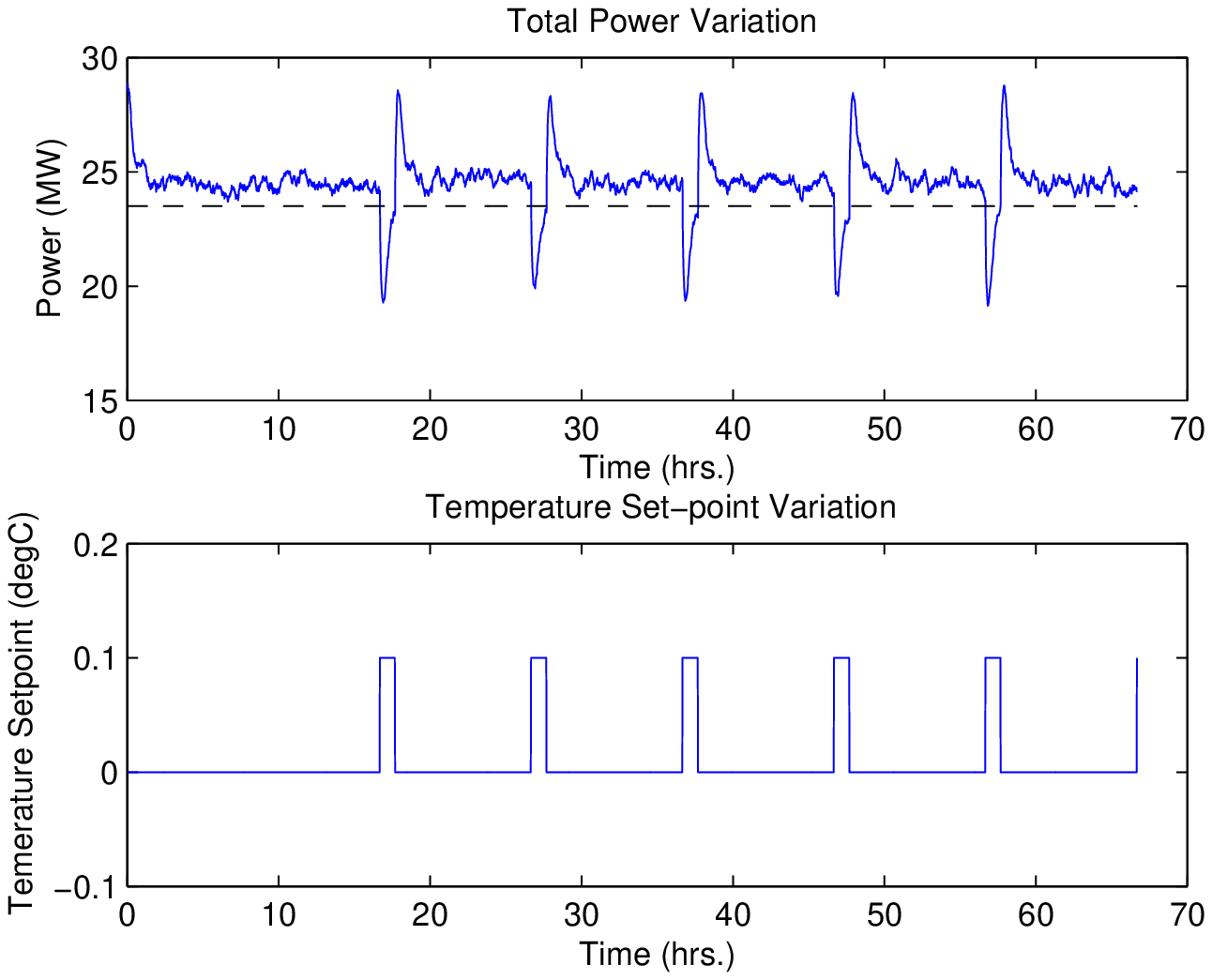, width=2.5in}\label{probing_power}}

\subfigure[Energy consumed (over nominal) in response to such pulses. (A negative value means energy is delivered.)]{\epsfig{file=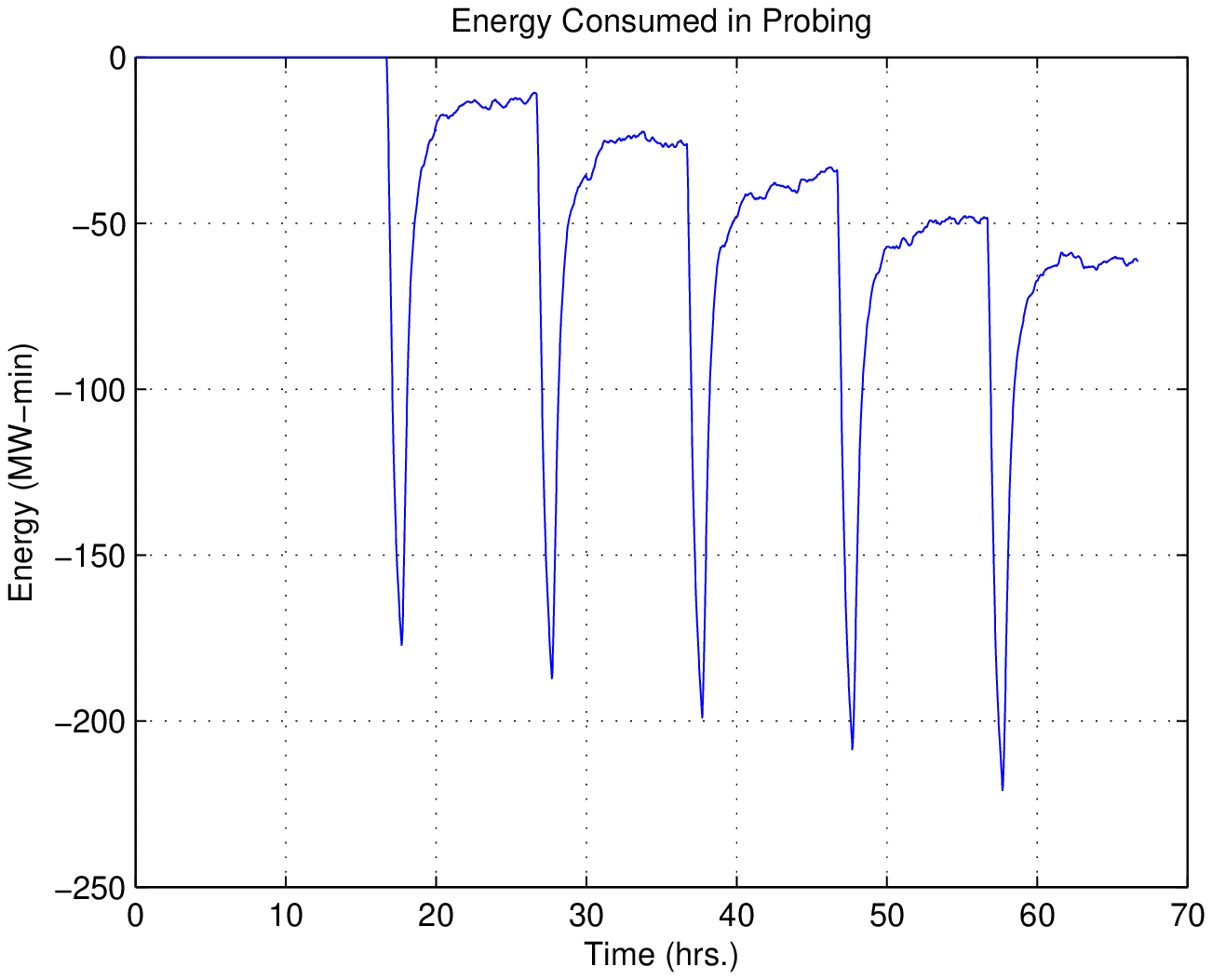, width=2.5in}\label{probing_energy}}
\caption{Energy consumption in the probing method.} \label{probing}
\end{figurehere}

Figure~\ref{probing_power} can be used to explain the probing
method. The temperature setpoint is increased and held at that value
for a short duration and then returned to its original value.  The
system is probed by short pulses spaced reasonably far from each other
in time. The energy delivered during such probing is monitored, with
Figure~\ref{probing_energy} providing an illustration. It can be seen
that the energy consumed, relative to the nominal consumption, is actually negative suggesting that energy is ``delivered'' by the loads when probed with positive
pulses. Knowing that over a certain duration a certain amount of
energy can be delivered by the loads, the pulses can be scheduled to
balance any under-generation. Similarly, over-generation can be
balanced using negative pulses.

\section{CONCLUSION}

In this paper we have analytically derived a transfer function
relating the change in aggregate power demand of a population of TCLs
to a change in thermostat setpoint applied to all TCLs in unison. We
have designed a linear quadratic regulator to enable the aggregate
power demand to track reference signals. This suggests the derived
aggregate response model could be used to allow load to track
fluctuations in renewable generation. The analysis has been based on
the assumptions that the TCL population is homogeneous and that the
noise level is insignificant. When such assumptions do not hold, we
propose a probing method that can be used to perform energy
balance. Further studies are required to incorporate the effects of
heterogeneity and noise into the model. Those extensions are important
for determining the damping coefficient.

Similar analysis can be used to establish the aggregate
characteristics of groups of plug-in electric vehicles, another
candidate for compensating the variability in renewable generation.

\section*{ACKNOWLEDGEMENT}

We thank Dr.~Michael Chertkov of Los Alamos National Laboratory, USA
for his support and useful insights throughout this work. We also
thank Prof.~Duncan Callaway for many helpful discussions.


\end{multicols}

\end{document}